\documentclass[letterpaper]{article} 
\usepackage{aaai2026}  
\usepackage{times}  
\usepackage{helvet}  
\usepackage{courier}  
\usepackage[hyphens]{url}  
\usepackage{graphicx} 
\usepackage{natbib}  
\usepackage{caption} 
\usepackage{algorithm}
\usepackage{algorithmic}

\usepackage{xcolor}
\usepackage{changepage}

\usepackage{framed}
\newenvironment{textttenv}{\ttfamily}{\par}
\definecolor{formalshade}{RGB}{242, 242, 242}
\newenvironment{formal}{%
  \MakeFramed{\advance\hsize-\width\FrameRestore}%
  \noindent\hspace{-4.55pt}
  \begin{adjustwidth}{4pt}{7pt}%
}
{%
  \end{adjustwidth}\endMakeFramed%
}
\usepackage{booktabs}
\usepackage{newfloat}
\usepackage{listings}
\DeclareCaptionStyle{ruled}{labelfont=normalfont,labelsep=colon,strut=off} 
\floatstyle{ruled}
\newfloat{listing}{tb}{lst}{}
\floatname{listing}{Listing}
\title{Whose Values? \\ Measuring the (Subjective) Expression of Basic Human Values in Social Media Posts}
\author {
    Ziv Epstein\textsuperscript{\rm 1},  Farnaz Jahanbakhsh\textsuperscript{\rm 2}, Tiziano Piccardi \textsuperscript{\rm 3}, Isabel Gallegos \textsuperscript{\rm 4}, Dora Zhao \textsuperscript{\rm 4}, Johan Ugander \textsuperscript{\rm 5}, Michael Bernstein \textsuperscript{\rm 4}
}
\affiliations {
    \textsuperscript{\rm 1}MIT, \textsuperscript{\rm 2}University of Michigan, \textsuperscript{\rm 3}John Hopkins University, \textsuperscript{\rm 4}Stanford University, \textsuperscript{\rm 5}Yale University\\
    zive@stanford.edu, farnaz@umich.edu, piccardi@jhu.edu, iogalle@stanford.edu, dorothyz@stanford.edu. johan.ugander@yale.edu, msb@cs.stanford.edu
}

\begin{document}

\maketitle

\begin{abstract}
The value alignment of sociotechnical systems has become a central debate, but progress depends on how human values are perceived in the content these systems surface and how such perceptions can be measured at scale.  Social media platforms are a prominent class of sociotechnical systems where algorithmic curation shapes exposure to value-laden content at scale. Large-language models offer new opportunities for measuring expressions of human values (e.g., humility or equality) in social media data, but value expressions can be subjective: different people will annotate the same post with different values. In this paper, we draw on the Schwartz value system as a broadly encompassing and theoretically grounded set of basic human values, and introduce a framework to \textit{personalize} the measurement of expressions of Schwartz values in social media posts at scale.  We collect 32,370 ground truth value expression annotations from N=1,079 people on 5,211 social media posts representative of real users' feeds. Due to the subjectivity of the task, we observe low levels of inter-rater agreement between people, and low agreement between human raters and LLM-based methods. In response, we construct a personalization architecture for classifying value expressions by learning from a small number of highly informative calibration annotations per user. In evaluation, we find that modeling these differences successfully yields value expression predictions that people agree with more than they agree with other people. These results contribute new methods and understanding for the measurement of human values in social media data.
\end{abstract}


\section{Introduction}
Social media provides a rich and complex substrate to understand patterns of human life, behavior, and discourse~\cite{lazer2009computational}.  The rise of large language models (LLMs) offers new opportunities for measuring a wide variety of constructs in social media data~\cite{ziems2024can,jia2024embedding}. One construct of interest for social media, growing out of the focus on value alignment of sociotechnical systems~\cite{russell2015research}, is \textit{human values}: the core beliefs that serve as guiding principles and in turn shape decisions, behavior, and perception~\cite{kolluri2025alexandria,bernstein2023embedding,stray2024building}.  Values expressed on social media, therefore, offer a reflection of discourse and sentiment, that in turn affect behavior. The possibility of measuring human values in social media posts opens new opportunities to build a better understanding of how discourse operates online, and to design the next generation of social media algorithms and platforms. Past work has shown how values and related moral priorities explain patterns of belief and behavior in domains as varied as vaccine hesitancy \cite{weinzierl2022hesitancy}, pro-environmental behavior \cite{schultz1998values}, social norms \cite{forbes2020social}, prosociality \cite{heilman2020personal},  human development \cite{inglehart2020modernization}, consumer behavior \cite{krystallis2012usefulness, stathopoulou2019effect}, and news story framing \cite{mokhberian2020moral}. Beyond understanding, the capability to measure values could also enable algorithmic objectives for feed-based ranking that support both individual and societal goals. In particular, there is growing interest in aligning ranking algorithms with societal values \cite{bail2022breaking, bernstein2023embedding,stray2024building,kolluri2025alexandria}, focusing on values such as democratic process \cite{jia2024embedding, piccardi2024social}, well-being \cite{stray2020aligning}, and downranking low-quality information \cite{epstein2020will}. 

\begin{figure}
\centering
\includegraphics[width=0.7\columnwidth]{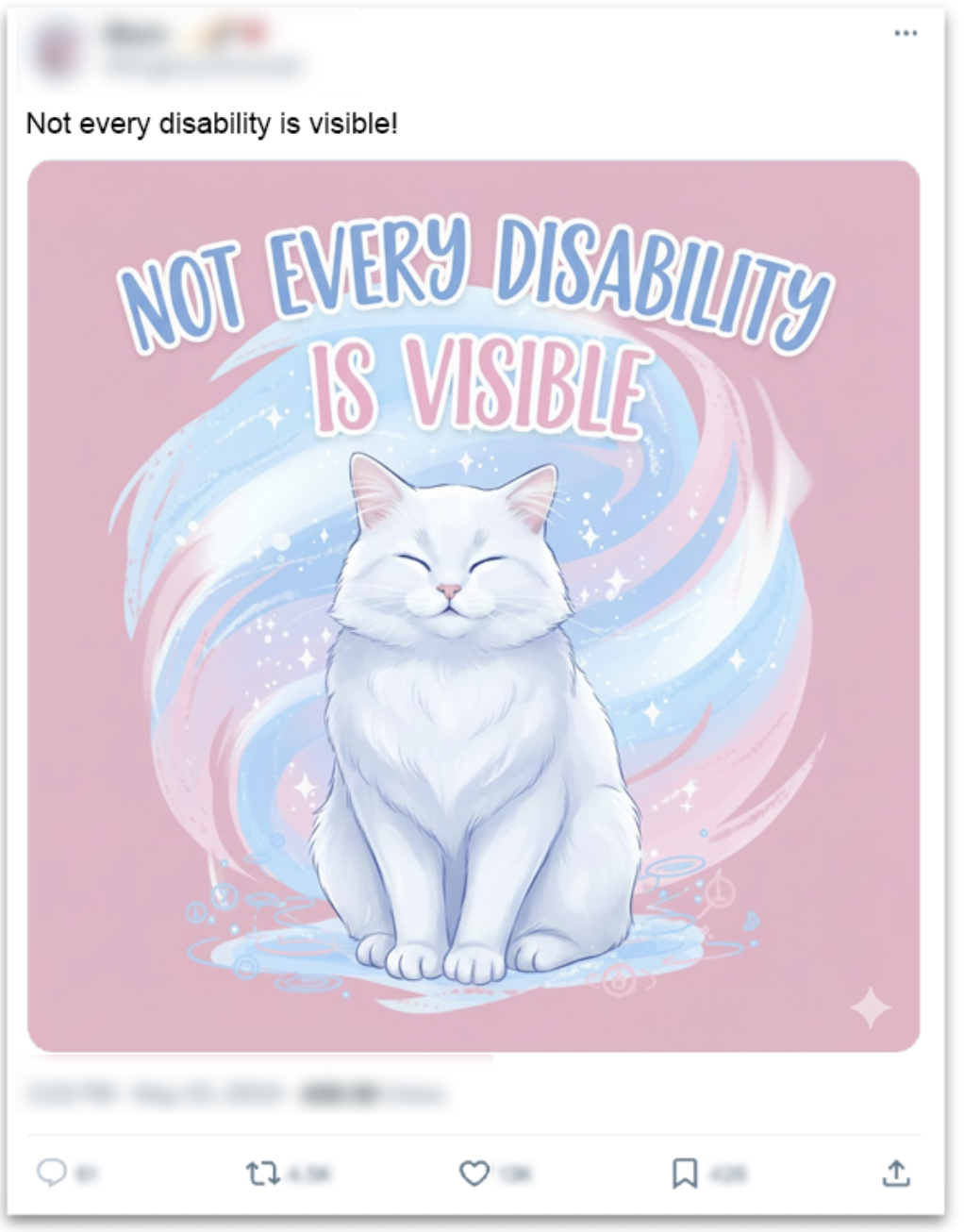}
\caption{An anonymized example social media post for value annotation. All annotators in our dataset agree that this post strongly expresses Universal Concern, and they all agree that the post does not express Tradition---but, annotators disagree substantially on whether the post expresses Humility. In this paper, we model these interpersonal differences for more accurate value classification.}
\label{fig:example}
\end{figure}

Before these opportunities can be realized, there remain both conceptual and practical challenges in operationalizing and measuring value expressions in social media. By \textit{value expressions}, we mean the extent to which a social media post expresses attributes associated with a given value. Identifying value expression can be straightforward in cases such as the post ``Not all disabilities are visible!'' (Figure~\ref{fig:example}), which many perceive as expressing the value of equality but not  tradition.

However, not in all cases are value expressions as straightforward to classify on social media.  The core issue is that value expression is a fundamentally \textit{subjective} construct, leading to different labels from different people~\cite{van2023differences,basile2021we,orlikowski2023ecological}.
For example, while the Figure~\ref{fig:example} post that ``Not every disability is visible!'' clearly is an expression of equality and clearly is not an expression of tradition, is it an expression of the value of humility? For some, the post strongly expresses a reminder to be humble about the conclusions we draw about other people's (lack of) visible disabilities. For others, their perception of that value expression is much weaker. Both interpretations are valid and shaped by an individual's background and how they both conceptualize these values and understand the social media post itself. As a result, we argue that there exists conceptual and methodological ambiguity about how value expression is operationalized. While existing work (e.g.,~\citet{qiu2022valuenet,van2023differences,borenstein2024investigating}) has developed classifiers for detecting basic human values, these models have presumed there exists in all cases one unambiguous correct answer to be learned.  But due to the possibility of divergent interpretations of value expressions, a traditional content-level ground truth set to train and evaluate models may obscure disagreements across individuals.

In this paper, we therefore argue for a paradigm shift toward value labeling architectures that directly acknowledge the subjectivity of these constructs. In particular, we first identify disagreements in perceptions of values in social media posts, and then develop personalized methods to identify an individual's perception of a value under a perspectivist framework. To do this, we collect a large annotated dataset where 1,079 people redundantly annotate values in 5,211 posts using a well-studied and theoretically grounded value system drawn from cross-cultural psychology --- the Schwartz value system~\cite{schwartz2022measuring,schwartz1992universals}. We evaluate to what extent large language models (LLMs) can identify the expression of values in these posts, comparing off-the-shelf LLM models with fine-tuned and personalized models, and both to individuals and groups of annotators.

On this task, we find that not only are there low levels of agreement between people, but also that an off-the-shelf LLM exhibits ever lower levels of agreement with people than people's levels of agreement with each other. To make sense of these findings, we draw on the theory of \textit{perspectivism} --- that different people will have different subjective experiences of the same situation \cite{soden2024evaluating, berger2016social} --- to suggest that people's divergent perceptions of both the posts and values are driving the low levels of human-human agreement we observe, as well as the lack of performance of the off-the-shelf (base) GPT model. 

With this idea in mind, we explicitly model interpersonal differences by combining two components: (1) fine-tuning a large language model to perform value inference, and (2) incorporating recommendation-system–inspired techniques that use a small, user-specific ``cold-start'' dataset to adapt predictions to individual users.
 
We find that our personalized approach results in a performant model that individuals agree with more than the rate at which they agree with other people (66\% relative improvement in spearman rank correlation ($\rho$)), or with the consensus vote (28\% relative improvement in $\rho$). 

We first contribute a demonstration that perceptions of value expression exhibit high levels of disagreement, which underscores the perils of relying on majority aggregated labels in this task. We then correspondingly contribute a theoretically-grounded personalization framework for calibrating predictions from a fine-tuned large language model to an individual based on their own responses. This results in predictions that align more closely with that person’s judgments than either other humans’ labels or few-shot–prompted LLM outputs.

\begin{figure}
\centering
\includegraphics[width=0.9\columnwidth]{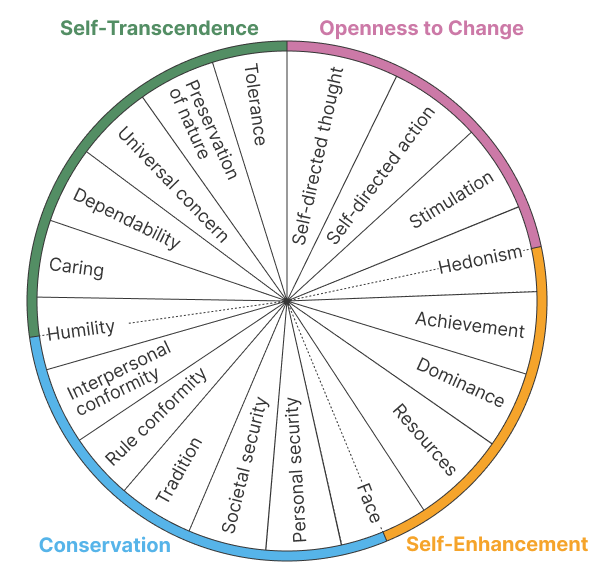}
\caption{The refined Schwartz value system. The system is hierarchically nested into a highest level: Outcomes for Self (right) vs Outcomes for Others (left), a high level: Self-Transcendence (top left, green), Conservation (bottom left, blue), Self-Enhancement (bottom right, orange) and Openness to Change (top right, red), and the 19 low-level values. Adapted from \cite{schwartz2012refining}}.
\label{fig:schwartz}
\end{figure}

\section{Related Work}

\subsection{Schwartz’s Theory of Basic Human Values}
To operationalize values, we employ Schwartz’s theory of Basic Human Values~\cite{schwartz2012refining}, a theory from Cultural Psychology for mapping individual values. Schwartz defined (basic) values as ``trans-situational goals, varying in importance, that serve as guiding principles in the life of a person or group'' and sought to identify a set of values that could be recognized in all societies. The original theory \cite{schwartz1992universals} identified 10 basic values from a 56-item  questionnaire administered to 9140 students and teachers across 20 countries. The refined theory identified 19 values that exist on a circular continuum with 57-item questionnaire administered to 6,059 participants across 10 countries (see Figure \ref{fig:schwartz}). A core assumption of Schwartz’s theory of Basic Human Values is that: ``The circular arrangement of values represents a continuum of related motivations, like the circular continuum of colors, rather than a set of discrete motivations''~\cite{davidov2008bringing}. This circumplex structure is also reflected in the hierarchical branching structure of the values, which starts at the highest level (Outcomes for Self vs Others), then branches further to four high-level values (Openness to Change, Self-Enhancement, Self-Transcendence, and Conservation), and then 19 low-level values. Values close to each other on the wheel are theorized to have similar and compatible underlying motivations, while values on opposite sides are theorized to be in tension.  While there are reasons to critique the Schwartz Value System (see e.g. \cite{chakroff2015discovering}), the theory ultimately offers conceptual precision, depth, and breadth, as well as offering more theoretical plausibility, analytical utility, and empirical grounding than political values \cite{goren2020values}. 

\subsection{Detecting and Using Schwartz Values}
\citet{qiu2022valuenet} create a dataset of human attitudes on over 20,000 text scenarios called \textsc{ValueNet} and a transformer-based regression model for dialogue tasks. \citet{kiesel2022identifying} create a dataset of the values behind arguments from four geographical cultures. \citet{ponizovskiy2020development} developed a dictionary of words for Schwartz values based on a corpus of Facebook updates, blog posts, essays, and book chapters.  \citet{boyd2015values} use free-response survey methods to compare individual values reconstructed from text with traditional survey instruments, showing advancements in understanding in both the structure and content of values \citet{van2023differences} use data from \citet{qiu2022valuenet} and \citet{kiesel2022identifying} to train a model for predicting value expression in Reddit posts, which they apply to a dataset of 11.4M comments from 19K users to construct value profiles of individual users. They find that dissimilarity of users' values is associated with the levels of disagreement between those users.
\citet{borenstein2024investigating} train value extraction models to classify both the presence and polarity of Schwartz values in reddit posts. They identify patterns like a strong negative stance towards conformity in subreddits such as r/Vegan and  r/AbolishTheMonarchy an overall correlation between tradition values and the conservativeness of U.S. states. \citet{shahid2025llms} use LLMs to re-write constructive comments on homophobic and Islamophobic threads, then uses human annotations to evaluate the values represented in these comments. They find that the comments rewritten by LLMs exhibited decreased Conservative values with increased prosocial values such as Benevolence and Universalism. 

This body of work uniformly makes the assumption that there exists an unambiguous ``ground truth'' value label, neglecting the inherent subjectivity of the task. In the next subsection, we discuss work that explores an alternative, perspectivist stance.

\subsection{Perspectivism}
The work highlighted above on automatically detecting Schwartz values offers an exciting lens on the values represented in text. However, in every case, these models assume a fixed meaning for both the content and the values, and a corresponding unambiguous ``objective'' value label in each case. This is in line with the longstanding paradigm in data labeling \cite{fleisig2024perspectivist} which seeks to collect ``ground truth'' labels \cite{nowak2010reliable, snow2008cheap} by aggregating labels from annotators. An alternative, the \textit{perspectivist} paradigm \cite{basile2021we, fleisig2024perspectivist, plank2022problem, mcdonald2019reliability, frenda2025perspectivist} argues that variation in annotator labels is not purely a source of error to be expunged, but rather is intrinsically meaningful. In this vein, \citet{haghighi2025ontologies} argues that the ontological dimensions underlying generative systems are under-recognized, and makes the case for perspectives that are plural (instead of universal), grounded (instead of abstract), lively (instead of fixed) and enacted (instead of diluted).

We argue that our task of identifying the expression of basic human values in social media posts fits this framework well, because the labels have no singularly correct ``ground truth'' label that everyone agrees with (they are plural) and are personally calibrated to each individual's behavior (they are grounded). 

\subsection{Personalization}
An alternative but complementary lens is personalization, which seeks to build models responsive to heterogeneous preferences. In the context of recommender systems, classic approaches involve matrix factorization \cite{koren2009matrix} and collaborative filtering \cite{resnick1994grouplens}. For social media algorithms, \citet{seth2008social} use a Bayesian user-model to learn personal recommender systems from social network data. \citet{lerman2006social} uses social filtering of activities as a way to personalize recommendations. See \citet{eg2023scoping} for a review on social media personalization. 

In the context of aligning LLMs, past work has explored aligning models to users' opinions \cite{hwang2023aligning, do2025aligning,suh2025language} and behaviors \cite{shaikh2025aligning}. This is often task specific, with advances in domains as varied as computer usage \cite{shaikh2025creating}, creative writing \cite{chung2025literarytaste} and user interface generation \cite{wu2025improving}. In addition, recent work has explored personalized reward models \cite{chen2024pad, jang2023personalized,li2024personalized, rame2023rewarded}. This is achieved by conditioning of text-based user profiles \cite{zhang2024guided, ryan2025synthesizeme}, using learned auxiliary predictors to steer towards group preferences \cite{zhao2023group}, and latent variable approaches \cite{poddar2024personalizing}. In this work, we adopt a cold start approach (e.g. identifying and using a small yet compact set of past behaviors) because of its flexibility and interpretability. 

\begin{figure*}[h]
\centering
\includegraphics[width=.99\linewidth]{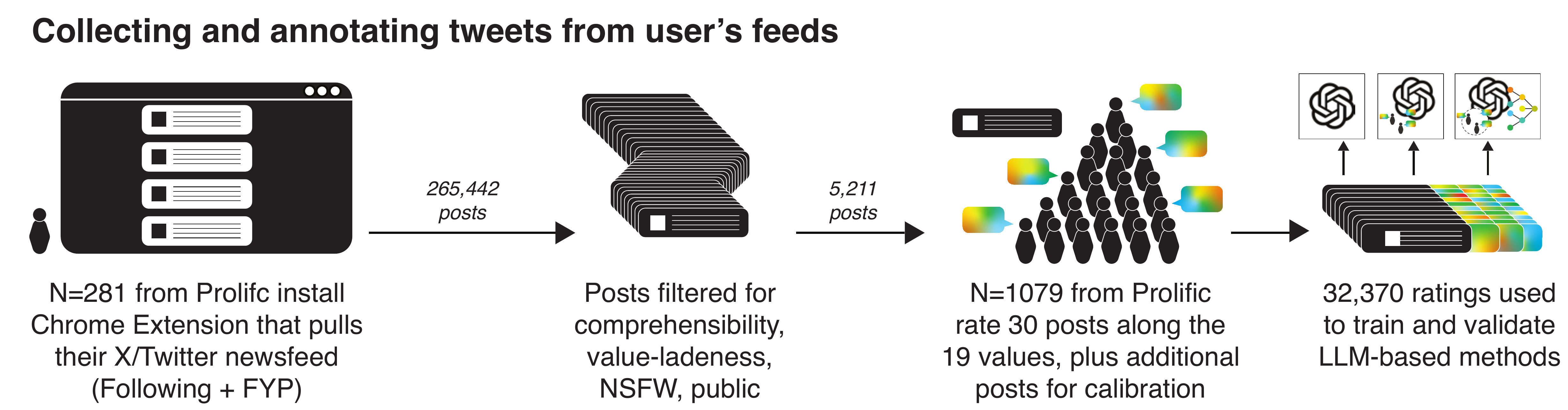}
\caption{Method for collecting and annotating tweets from user’s feeds}
\label{fig:flowfig}
\end{figure*}
\section{Methods}

In this section, we first report our methods for curating a dataset of social media posts. Next, we discuss our methods for annotating this dataset for values. Finally, we provide details our approach to modeling and evaluating human and AI-based value expressions. 
\subsection{Curating a Dataset of Social Media Posts}

\subsubsection{Recruitment}
We recruited a sample of social media users (N=281) quota-matched to be a representative sample of the US population on age, ethnicity, gender and partisanship using Prolific in May 2024. Participants were all from the United States and were required to be 18 years old or older. We were unable to meet the quota for 17/109 participants in the age range 55-100 so we upsample posts from participants in this age range by 18.4\% when constructing our silverset of social media posts. Within this sample, the median age was 43 (min=19, max=80), 58.36\% White/Caucasian, 51.2\% female and 30.6\% Democrat, 28.8\%  Republican and 40.6\% Independent. Participants were paid $\sim\$15.42$ an hour and took approximately 7 minutes to complete the study. 

\subsubsection{Feed Collection}
After they consented to the research, participants were asked to install a Chrome Extension that downloaded posts from their X (Twitter) feeds, both their algorithmically curated For You page (FYP) and their in-network Following feeds simultaneously \cite{piccardi2024reranking}. We collect and annotate posts from both the Following feed and FYP to account for any systematic differences that may exist between them, and to pave the way for work that uses both to measure algorithmic amplification (see Discussion).

When a browser opens the X page, tweets are fed to the browser client in ``batches'': a single API request returns a set of tweets at a time. The Chrome Extension aims to download 16 batches of tweets: 7 from the FYP, and 9 from the Following page.
Each batch for FYP consists of approximately 30 tweets, whereas the Following feed returns a variable number of tweets up to 100.

From the 281 participants, we collected 265,442 posts, with an average of 207 posts from the FYP and 776 from the Following page per user. Among these 265,442 posts, there are 151,740 posts with unique, non-empty text in English (using Google's Chromium language detection tool \texttt{cld2}). We also screen out private posts (n=6,025; 3.97\%), deleted posts (n=2,004; 1.32\%) and/or reply tweets that are replying to posts not displayed in the user's feed (n=7,581; 4.99\%) to focus only on the 142,652 public posts we can render with proper context. 

\subsubsection{Subsampling feed posts to create an annotation dataset}
We next prepare these posts for human annotation, since some posts are either not safe for work (NSFW) to show to annotators or incomprehensible to most annotators. We filter the posts for comprehensibility (grounded in the constructs of readability, coherence, spam behavior, and context required for understanding) and NSFW (to comply with our institution's IRB for data annotation) using an automated GPT-based pipeline, (see Table~\ref{tab:nsfw} for full text and the Appendix for an evaluation of this model) with \texttt{GPT-4o}, temperature=1.0 and seed=0. We filter out any posts that score less than 3 on a comprehensibility scale, resulting in a dataset of 89,383 (62.6\%) posts. 
We then filter out 1,937 posts with NSFW content, resulting in a final dataset of 87,446 posts.

The occurrence rate of different values in social media posts varies, as do the numbers of posts per person across both Following and FYP feeds. Therefore, uniform sampling procedure may yield a sample that underrepresents rarer values and overrepresents certain users for annotation. To construct a final set of posts for human annotation, we stratify on user and post source (FYP vs following) as well as on a preliminary LLM-based screening measure of value expression, designed to err on the side of inclusion and capture as many plausibly value-expressing posts as possible.

In particular, we perform a preliminary rough classification of these posts for all 19 of the Basic Human Values \cite{schwartz2012refining} on a 7-point Likert scale [0-6] using another automated GPT-based pipeline with the prompt that rates all 19 values jointly each on a 0-6 scale based on a short description, with the prompt defined in Table~\ref{tab:values} with \texttt{GPT-4o}, temperature=1.0 and seed=0 (see Appendix for an evaluation of this method). This allows us to upsample posts that may reflect rarer values. For this upsampling, we say a given post reflects a given value if the GPT annotation pipeline scored it a 4 or above. We then sample a set of 5,227 posts for human annotation by stratifying across users, post source, and value expression. In particular, for each user and each value, we sample a tweet from that user's feed that reflects that value. We sample FYP and Following posts with equal probability, but for a given user, if only one type of post exists, we return that tweet. Because we did not meet our quota for participants over the age of 55 (by 17 participants), we randomly sample (without replacement) 17 of the 76 participants over the age of 55 who completed the survey and repeat this process for each of them to ensure representation of posts from users over the age of 55.  Of the 5,227, 3,182 came from users' Following feeds (60.8\%) and 2,045 (39.1\%) came from users' FYP. This process yielded a dataset of 5,227 posts that we can use to study human value annotation and to predict the values expressed in those posts. 

\subsection{Annotating Social Media Posts for Schwartz Values}
In this section, we describe our ground truth data annotation method.

\subsubsection{Participants.}
In total, 1,276 participants began the survey. We filtered out N=153 participants who failed two simple attention checks (see Appendix) or who did not finish the study. N=1,123 completed the study. Our final dataset included ratings from N=1,079 participants who have a full set of ratings for all 30 posts assigned to them and within this sample, the median age was 45 (min=18, max=85), 61.2\% White/Caucasian, 50.8\% female and 29.6\% of participants lean Democrat vs 28.63\%  lean Republican.  Participants were paid \$15.00 an hour and took approximately 60 minutes to complete the study. 

\subsubsection{Procedure }  
Each participant provided value expressions ratings for 30 posts. The posts for each participant were randomly sampled from the dataset of 5,227 posts sourced as described above (each post was rated by a median of 6 people). Given this sampling procedure, we retained ratings of 5,211 posts (since posts are sampled with replacement for each participant). 

We used a recursive labeling where raters could traverse the value tree as defined in Figure~\ref{fig:tree} (top) devised to elicit value ratings for all 19 values in an efficient manner. The branching method leverages the tree structure of the Schwartz Theory of Basic Values, starting with the high-level values, and traversing down only those branches whose parent values they had marked as expressed in the post. 
 
Following this primary task, participants completed a custom developed Value Calibration Questionnaire (VCQ) of 25 additional ratings on a shared set of questions (see Table~\ref{tab:vcq} and the Personalization framework in the Methods sections for a description of how these questions were sourced). Finally, participants answered a number of demographic and political questions. 

To familiarize them with the task and to determine eligibility for proceeding to the main labeling task, participants first completed a short training for four synthetic posts followed by a gating task. The training introduced annotators to the meaning of the Schwartz values, while the gating task assessed their understanding. These training posts were created to unambiguously express particular values, and participants could not proceed until they checked the right answer: if they answered incorrectly, a message appeared providing the correct answer and asking them to try again. 

After training, they proceeded to the gating phase, where they rated four posts that, based on a pre-study, either unambiguously contained or did not contain the values of \textit{Self-directed action} or \textit{Face}. Participants who got one or fewer of the four correct were not allowed to continue.  We took care in selecting the questions for both pre-tasks and in setting the eligibility threshold for the gating task to ensure a shared baseline understanding. See Figure~A\ref{fig:tree} Bottom for screenshots for the annotation of one of the gating posts. 

After completing basic attention screeners and training, participants who successfully passed the gating phase proceeded to rate 30 posts drawn at random from the larger set of 5,227 posts for value expression using the branching method discussed above. When presenting a tweet to our annotators for value annotation, to convey context, the text of a given tweet (tweet\_1) made in reply to or quoting another tweet (tweet\_2) was represented as ``tweet\_2 REPLY TO: tweet\_1'' and ``tweet\_2 QUOTED: tweet\_1'' respectively. The annotation dataset we create is available for researchers on a case by case basis to preserve the privacy of the participants. It is stored securely in a de-identified manner.

\subsection{Modeling Basic Human Values}
In all instances, we compare the predicted value ratings from a range of algorithmic models to ratings from individuals or groups of annotators on a fixed holdout test set of 1496 (28.2\%) of posts, which allows us to evaluate the finetuned and personalized models for unseen posts. The remaining 3715 posts are used for training models. We note that we hold out at the \textit{post}-level, letting train on a subset of the ratings for a given participant and test on their ratings from a different subset of posts. 

To improve performance beyond zero-shot GPT-4o, we perform finetuning on high-consensus value labels. To produce this training set, we randomly select 1000 posts for which we have more than 6 ratings per post from our dataset of 5,211 posts annotated by 1,079 participants across the 19 values. Then, for each post, we compute the Spearman rank correlation in value ratings for each pair of raters and average these correlation scores into an aggregate consensus score for each post. We then use the 600 highest consensus annotations of these posts (e.g., average value score rounded to the nearest integer, 11.5\% of posts) as labels to fine-tune GPT-4o via the OpenAI API, producing a model that estimates the consensus label of the annotators. The fine-tuned model reached a final error of 0.0722 after 600 steps. 

\subsection{Personalization framework}
\begin{figure}[t]
\centering
\includegraphics[width=.99\linewidth]{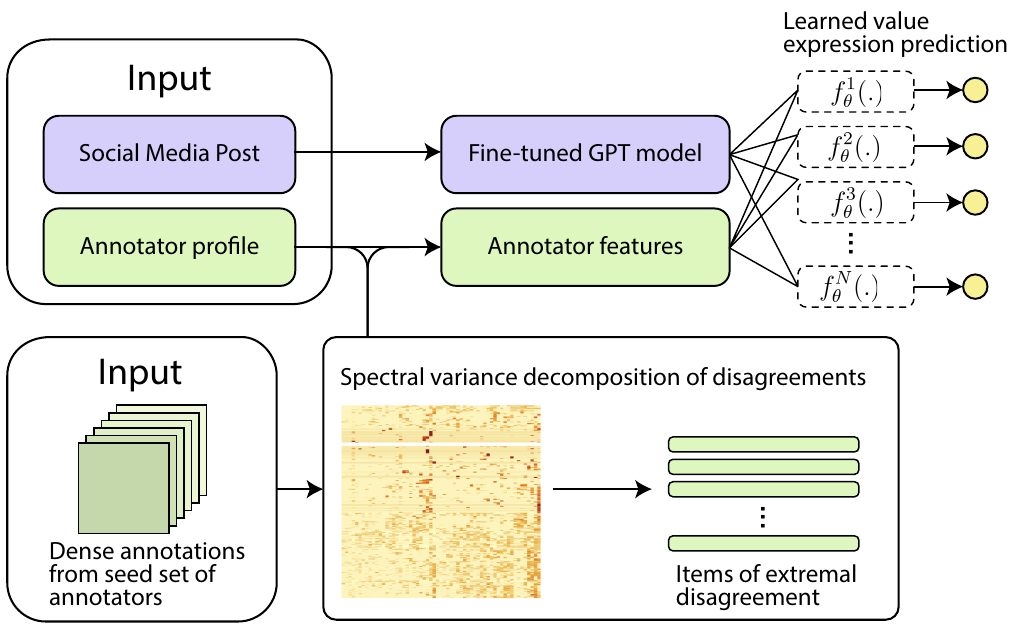}
\caption{Our personalization framework uses a bespoke set of annotator features to calibrate predictions from the fine-tuned LLM. }
\label{fig:arch}
\end{figure}

To capture the subjective nature of value expression annotation, we combine the consensus fine-tuned model with a personalized calibration model to predict the values that a given \textit{individual} would perceive as expressed in a given post. We do so by fitting a series of models (one for each value) trained on the ratings of 3000 posts (with 1496 heldout for evaluation) stratified by number of ratings to predict a given individual’s value annotation of a given post using (a)~the consensus predictions from our fine-tuned model as well as (b)~that individual's responses to the calibration questions designed to explain variation in rater disagreement (see Figure~\ref{fig:arch}). First, we describe how the calibration questions were developed, then we discuss the modeling procedure. 

To identify features for modeling that are highly predictive of interpersonal differences, we identify post-value pairs that explain maximal variation in value expression annotation disagreements in a pre-study. In this pre-study, N=51 participants rated the same set of 30 posts on all 19 values using the same procedure as the primary annotation study. But unlike that process, which randomly sampled posts for annotation by each participant, here every participant annotated all 30 posts, resulting in a dense tensor (participant$\times$value$\times$post).  We then transform the data into columns for each participant  (N=51) and rows for each post/value pair ($19\times30=570$). 

We de-mean this matrix row-wise and conduct principal component analysis (PCA) to compute bases of maximal variation (``eigenraters''). The first 25 eigenraters explain 86\% of variation in disagreements across raters. For each of these eigenraters, we find the extremal post/value pair with the highest absolute value and add it to our set of questions, which we dub the Value Calibration Questionnaire (VCQ) to use for the main annotations study to solve the cold start problem. Each of the 25 questions presents a post and asked the participants to rate the extent to which a specific value was reflected in the post, phrased as ``To what extent does this post reflect \{value\}?''  The full set is shown in Table~\ref{tab:vcq}. 

Given this data-driven approach to decomposing variance in value disagreements, post/value pairs are selected for inclusion because they are extremal within an eigenrater, not because they exhibit coverage of the items or values. We therefore note the appearance of both duplicate posts that sparked disagreement across different values (e.g., indices 3/16/22, 5/17, 7/18, 10/11, 2/25, and 13/23 in Table~\ref{tab:vcq}) and corresponding duplicate values that sparked disagreement across different posts (\textit{Caring}, \textit{Tradition}, \textit{Equality}, \textit{Achievement}, and \textit{Independent thinking}). This also means certain values do not appear in the VCQ such as \textit{Power}, \textit{Societal security}, \textit{Respect}, \textit{Humility}, \textit{Responsibility}, and \textit{Connection to nature}, which is one limitation of this approach (see Discussion). 

Our personalized framework does not directly model individuals---it instead uses the 25 calibration questions as individual-level features to solve the cold start problem (see  Figure~\ref{fig:arch}).  For each post–user pair, these models predict that user's 0–6 rating for a given value V using a random forest (e.g., one random forest model for each of the 19 values).  Each model was trained in R using the \texttt{randomForest} package and default hyperparameters (500 trees). Figure~\ref{fig:rfflow_sup} shows the total decrease in node impurities (variable importance) of each of the 19 random forest models, revealing how for a given value $v$ the random forest predicting $v$ primarily uses post-level the fine-tuned GPT prediction for value $v$ (shown on the top diagonal).

\subsection{Evaluation Approach}\label{eval}
In settings where there is high levels of subjective variation exists in the labels, a traditional content-level ground truth set is not an appropriate measure and instead evaluations should rely on measure agreement with disaggregated individual labels \cite{gordon2022jury, gordon2021disagreement}. For our evaluation metric, we chose Spearman rank correlation for several reasons.  For one, a correlation metric allows us to compare the \textit{relative} expression across all 19 low-level Schwartz values, which follows the relational nature of the value system itself. Further, a correlation computation across values is inherently scale-invariant, allowing us to account for individual level differences in how the scale is used. We also chose Spearman rank correlation because the rank ordering captures the ordinal prioritization and relational nature of values consistent with the circumplex theory of the Schwartz theory of Basic Human Values. However, future work might explore extensions that explicitly use the circular geometry of the circumplex directly (see Discussion).  We note that our results are qualitatively similar when using Pearson correlation instead of Spearman. In addition, a correlation metric handles the ordinal nature of the value annotations, which exist on a 7-point Likert scale. 

We note that this choice of Spearman rank correlation is an extremely \textit{conservative} measure, providing a rigorous lower bound on the actual levels of agreement. This is because the structure of the rank correlation, computed between two 19-dimensional values taking on values [0...6] with much density on 0, is quite sensitive to small adjustments in values, as incrementing and decrementing one value can cause it to skip up or down many other values in the rank order. Furthermore, in cases of posts with very little value expression (e.g. a vector of 18 zeros and 1 one), calculating Spearman rank within another very similar vector (e.g. a vector of 18 zeros and 1 one in another position) will result in a rank correlation of -0.05. And in the limit, calculating Spearman rank between two vectors of all zeros results in NA (because there is no variance) and is excluded from analysis, even though these cases represent perfect agreement. These reasons indicate that this extremely \textit{conservative} measure, should not be considered in absolute numeric terms of model quality, but rather relative terms to establish that people agree with our LLM-based value annotation model more than they agree with a random other person, or consensus of groups of increasing size. 

For a more interpretable evaluation metric, we also report results on MAE over values. In particular, we compute the consensus label of a holdout set of 1,496 posts by taking the mean rating rounded to the nearest integer for each value. Then for each post, we report the mean absolute error (MAE) between the LLM-generated label and the consensus label as LLM-Consensus MAE for each value. We then compare this error to the mean average distance that an individual human annotator has with the overall consensus label (Human-Consensus MAE). To be more precise, for each of the $k$ annotators who labeled a value $\mathcal{V}$ on a given post, we compute the MAE between the label they assigned to the value and the mean labels of the remaining $k-1$ annotators as follows: $\frac{1}{k}\sum_{i=0}^{k} | \mathcal{V}_i - \frac{1}{k - 1}\sum_{j=0, j \neq i}^{n}\mathcal{V}_j |$.

\section{Results}

\subsection{To what extent do people agree?}

\begin{figure}[t]
\centering
\includegraphics[width=.99\linewidth]{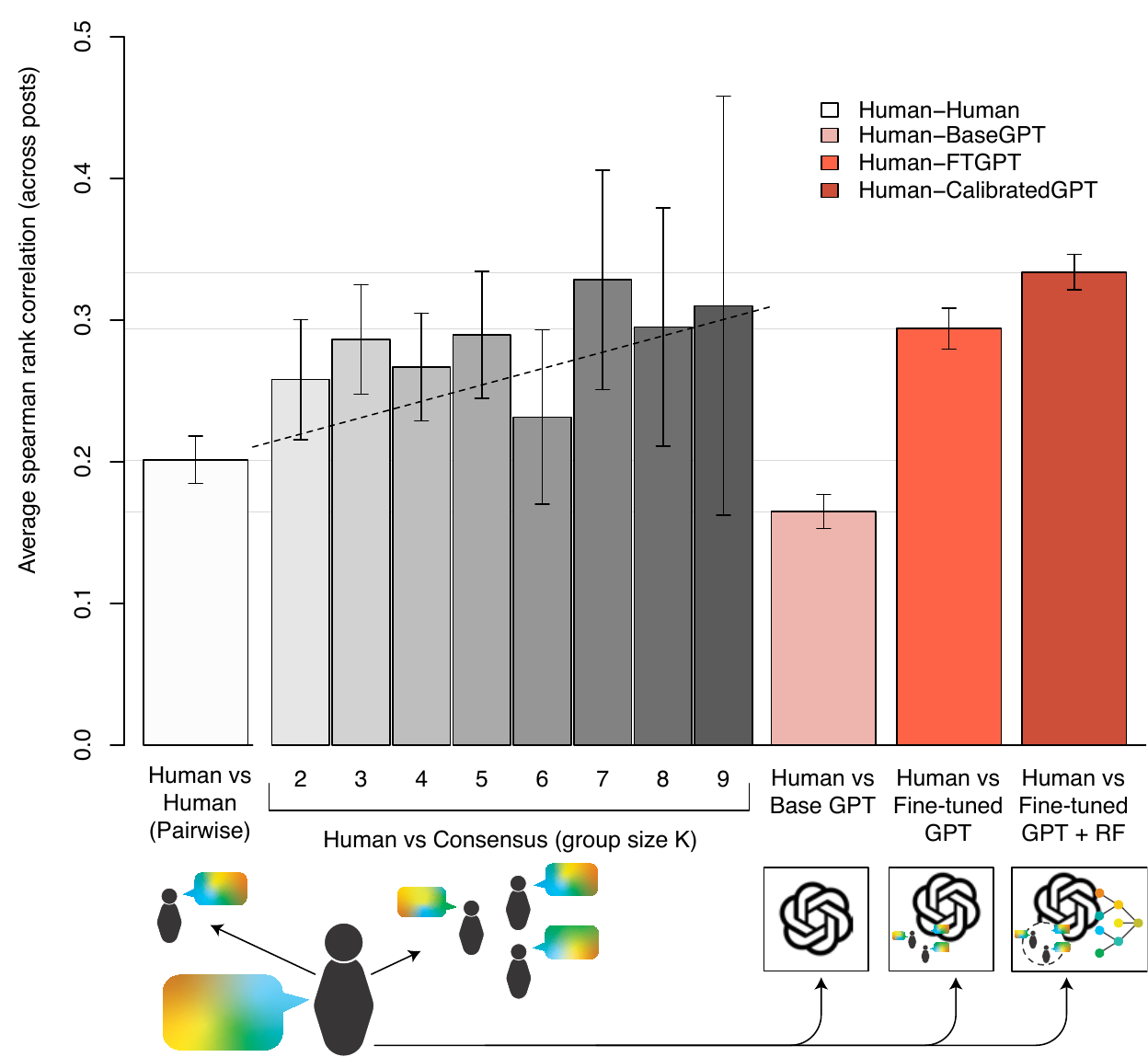}
\caption{Average Spearman rank correlation (across posts) comparing individual value annotations to 1) other individuals (gray, far left),  2) groups of increasing group size (light gray to dark gray, middle), 3) zero-shot GPT-4o (light red, right), 4) fine-tuned GPT-4o (red, right), and 5) personalized random forest (dark red, right).}
\label{fig:mod_res}
\end{figure}

 We observe low levels of agreement in value annotations among raters, with the average Spearman rank correlation across posts of 0.201 (Figure~\ref{fig:mod_res}, white bar).  When comparing individual annotations to the consensus annotation for that post (i.e., the average of all other participants' ratings, rounded to the nearest integer), we observe an overall average Spearman rank correlation of 0.260 and an overall positive relationship with the number of raters (p=0.037), indicating a 29\% relative improvement in the correlation in human-to-consensus labels compared to human-to-human labels. This suggests that despite the low levels of agreement, some levels of consensus are achieved when aggregating across individuals, particularly more than 3 or 4 people. 
 
 We also find subjectivity in value ratings, which helps explain the sources of disagreements we identify above. To test this claim we predict individual's value annotations across the 19 values by regressing, for each value, that particular value against each of the individual values that that individual holds, as well as demographics such as age and partisanship. The resulting regression coefficients are shown in the bottom of Figure~\ref{fig:valueheatmap}. We find that for 8/19 of the values, the extent to which an individual holds that value is predictive of the likelihood that they would annotate a given post with that value (\textit{Dominance}, \textit{Resources}, \textit{Achievement}, \textit{Self-directed thoughts}, \textit{Self-directed actions}, \textit{Stimulation}, \textit{Personal security}, and \textit{Societal security}), which suggests people may project their own values into the perception of those values onto novel situations.  We observe some partisan differences, where Republicans are more likely to rate a given post as reflecting the values of \textit{Hedonism}, \textit{Self-directed thoughts}, \textit{Self-directed actions}, \textit{Stimulation}, \textit{Humility}, and \textit{Nature}, which may map onto cultural-political divides. Within our sample, we do not detect age effects, except for the value of \textit{Interpersonal conformity}. 

\subsection{Predicting Basic Human Values with LLMs}
Next, we compare rates of agreement among people with those between people and LLM-based methods. First, we find that people agree with the zero-shot GPT-4o with an average Spearman rank correlation of 0.165 (light red in Figure~\ref{fig:mod_res}), which is a 18\% relative decrease in the correlation with which they agree with the annotations from a random other person ($\bar{\rho}=0.201$, white bar in Figure~\ref{fig:mod_res}, $p<0.001$).

By fine-tuning GPT-4o on the consensus annotations from the highest consensus posts, we construct an AI model that generates value annotations with which people agree more than they agree with a random other person ($p<0.001$), with an average Spearman rank correlation of 0.294. This improvement reflects a 46\% relative improvement in correlation compared to the human-to-human correlation. Comparing the fine-tuned GPT-4o to these wisdom of the crowds estimates, we see a rate of agreement similar to taking the average (consensus) rating of 8 or more other people (medium red in Figure~\ref{fig:mod_res}).  

When aggregating across values, we find the average mean absolute error (MAE) between the fine-tuned GPT-4o and human consensus (crowd) is $0.763\pm0.037$ compared to an average MAE between human consensus and individual raters of $1.113\pm 0.0743$ (see Table~1). This result suggests that the LLM is predicting the consensus better, since its distance from the consensus is less than from a random individual annotator.

\begin{table}[h]
    \centering
    \begin{tabular}{lrrr}
    \toprule
        Value &  Human-Crowd  & LLM-Crowd  \\
          &   MAE & MAE \\
    \midrule
Humility & $ 1.620\pm0.035$ & $0.971 \pm0.024$\\
Caring & $1.622\pm0.039$ & $1.027\pm0.025$\\
Dependability & $1.512\pm0.037$ & $ 0.917 \pm0.025 $\\
Universal concern & $1.389\pm0.039$ & $0.865\pm0.025$\\
Hedonism & $1.325\pm 0.037$ & $0.927\pm0.026$\\
Face & $1.208\pm0.037$ & $0.918\pm0.024$\\
Tolerance & $1.208\pm0.036$ & $0.740\pm0.023$\\
Societal security & $1.206\pm 0.036 $ & $0.779\pm0.024$\\
Rule conformity & $1.175\pm0.037$ & $0.737\pm0.024$\\
Tradition & $1.169\pm0.035$ & $0.777\pm0.024 $\\
Self-directed actions & $1.118\pm0.038 $ & $ 0.822\pm0.025$\\
Self-directed thoughts & $1.104\pm0.038$ & $0.800\pm0.024$\\
Interpersonal conformity & $1.048\pm0.035$ & $0.699\pm0.022$\\
Achievement & $1.005\pm0.037$ & $0.749\pm0.022$\\
Stimulation & $0.959\pm0.035$ & $0.729\pm0.024$\\
Preservation of nature & $0.738\pm0.031$ & $0.497\pm0.022$\\
Resources & $0.694\pm0.030$ & $0.599\pm0.020$\\
Dominance & $ 0.650\pm0.029 $ & $0.566\pm0.018$\\
Personal security & $0.414\pm0.026$ & $0.395\pm0.017$\\
\midrule
Overall & $1.113\pm0.0743$ & $0.763\pm0.037$\\
\bottomrule
    \end{tabular}
    \caption{Mean Absolute Error (MAE) between human value expression annotations and consensus (crowd) aggregations (center) and MAE between the fine-tuned gpt-4o model and consensus (crowd) aggregations (center) on a holdout set of posts (N=1,496)}
    \label{tab:mae}
\end{table}

\subsection{Personalization Results}
While fine-tuning an LLM on high-consensus labels yields a model that people agree with more than they agree with other people, the inherent subjectivity of the task underscores the point that any single model will not be performant for everyone. Therefore, next we train a \textit{perspectivist} model that attempts to model an individual's subjective perception of value expression. In particular, the personalized calibration model uses a series of random forest models to predict the values that a given individual would see reflected in a given post. Using only an individual’s calibration ratings and the target post’s aggregate GPT-predicted values as inputs, the random forests achieves an average Spearman rank correlation of 0.334 with human ratings, a rate of agreement surpassing the wisdom of the crowds ($p<0.001$, dark red in Figure 4). This most performant model reflects a 66\% relative improvement in correlation vs. the human-to-human baseline. This model also reflects a 28\% relative improvement in correlation vs. the more difficult human-to-consensus condition. We find that for each value, the random forest models typically use the underling GPT consensus prediction of that value as the most important variable (see Figure ~\ref{fig:rfflow_sup}) but also make use of the calibration questions from Table~\ref{tab:vcq}.

These results (Table~\ref{tab:results-corr}) support the claim that people agree with the LLM-based method at a level at or above the extent to which they agree with other people, for both other individuals as well as wisdom of the crowds.

\begin{table}[t]
\centering
\small
\caption{Correlation results and relative improvement vs.\ human–human baseline.}
\label{tab:results-corr}
\begin{tabular}{lrr}
\hline
\textbf{Condition} & \boldmath$\rho$ & \textbf{\%$\Delta$ vs.\ H--H} \\
\hline
Human vs.\ Base GPT         & 0.165 & -18\% \\
Human vs.\ Human            & 0.201 & --  \\
Human vs.\ Consensus        & 0.260 & +29\% \\
Human vs.\ Fine-tuned GPT   & 0.294 & +46\% \\
Human vs.\ Personalized GPT & 0.334 & +66\% \\
\hline
\end{tabular}
\end{table}

\subsection{What Drives Personalization Improvements?}

\subsubsection{Adjusting differentially subjective values} In Table~\ref{tab:mae}, we see that  the most overall subjective values in our sample were humility, caring, and dependability, while some of the most agreed upon values were resources, dominance, and personal security. To understand which values are being adjusted and how that effects personalization improvements, for a given value $v$, we compute the average absolute difference ($\Delta$) between the predictions of the fine-tuned consensus and the personalized models for value $v$. We find that the top adjusting values are humility ($\Delta=0.771$), caring ($\Delta=0.655$) and dependability ($\Delta=0.655$), and the least adjusting values are personal security ($\delta=0.344$) and preservation of nature ($\Delta=0.429$) with an overall value-wise spearman rank correlation of R=0.833. This indicates that personalization improves agreement by differentially adjusting subjective values where perceptions differ. 

\subsubsection{Personalized to whom?}
For each participant, we compute the average spearman rank correlation between that participant's ratings of a given post and the model predictions and then compute the difference in average spearman rank correlation between the consensus model and the personalized. In other words, we compute a participant-level measure of the delta between the red bar in Figure~\ref{fig:mod_res} and the dark red bar in Figure~\ref{fig:mod_res}. We then use a linear regression to compute the association between this delta (e.g. performance gain due to personalization) with individual differences. We find that personalization gains are relatively higher for men ($p=0.0102$), Republicans ($p=0.0297$) and no effect across ages ($p=0.824$). 

When comparing posts sourced from the For You page versus the Following page, we see no differences in either personalization gains ($p=0.709$) or individuals average level of disagreement with the consensus annotation ($p=0.8$).

\section{Discussion}
In this paper, we introduce and validate a framework for measuring the expression of basic human values on social media paradigmatic of a shift towards perspectivism. From a methodological perspective, we note that the large variation in value perceptions across individuals suggests that while indeed there is  a degree of shared understanding, there is no singular ground truth in every case: the low levels of agreement in value annotations we observe appear even after training and gating to encourage shared understanding. 

To acknowledge the inherent subjectivity of the task, we introduce a specialized framework for using LLMs for annotating subjective social science constructs. With this framework, we showcase a personalized LLM-based model that people agree with more than the base model, other people, or even consensus annotations. This approach represents a perspectivist approach \cite{fleisig2024perspectivist} to modeling subjective constructs, which we believe is critical in the age of ontologically monolithic LLMs \cite{haghighi2025ontologies}.

We note that in this paper, we were primarily focused on arguing for a paradigm shift towards personalization, and our personalization framework using random forests and a spectral decomposition of variations in disagreement is intended to demonstrate one possible instantiation that yields improvements beyond consensus annotations. 

The personalization framework presented here therefore represents one exemplary approach as a proof-of-concept but future work is needed to evaluate this approach more deeply and contrast it to a more comprehensive design space. For example, as discussed above, the method of identifying highly predictive features to solve the cold start problem does not include certain values and thus requires future work to understand how items identified generalize from one sample of raters (and items) to another and develop alternative methods for solving the cold start problem efficiently. 

In addition, we note that we only considered perspectivist models for the Schwartz value system. Future work should explore adapting this approach to other value systems that may be well-suited for the particular domain (such as the social media context). Depending on the domain, this might include other existing theories of values such as the World Values Survey or Moral Foundations Theory, or methods for deriving bottom-up theories of values via attentional probes \cite{klingefjord2024human}, or inverse reinforcement learning \cite{oliveira2023culturally}, or combinatorially combining ``moral molecules'' \cite{curry2022moral}. In addition, we note that the metric we used, Spearman rank correlaiton, does not explicitley capture the circualar geometry of the Shwartz value system and future work could explore bespoke measures that capture angular distance more precisely. 

Another limitation is that this study recruited Americans to annotate X posts. How this generalizes beyond the US content and to other social media platforms remains important future work. We also note we used two preliminary GPT‑based filters (comprehensibility, NSFW) in addition to a GPT‑based upsampling by predicted value to identify relevant posts that conformed to our IRB. Future work should investigate how value labeling extends to incomprehensible, value-sparse and NSFW posts. 

There are several both epistemic and practical implications of this shift towards personalized values that are important to keep in mind.  Epistemically, this underscores the need for further research into the subjectivity of a variety of social constructs (such as ``helpfulness''  or ``toxicity''), particularly when they are measured and deployed in high stakes sociotechnical contexts. These constructs may operate differently for different people, and therefore these systems may inadvertently exhibit bias based on divergent interpretations. 

One practical implication of this work is for social media feed design. If a platform can carefully understand how a user perceives values in content, it could tune that user's feed to highlight posts that resonate with their values. This opens the door to new feed algorithms that provide increased user control and alignment. In addition, this shift also provides a more precise way to audit what values are currently being amplified by existing feed ranking algorithms.  

\section{Ethical Considerations}
Our approach imagines a world in which value expressions can be readily assessed and measured. This may open the door to new forms of persuasion and microtargeting based on value-based rhetoric. This could be mitigated by mandated disclosure for advertisers and opt-out mechanisms for consumers. It could also include labels for value-laden language, to help consumers identify and contextualize value aligned or value misaligned content. 

This may also enable new forms of ranking algorithms that rank by values, which represents a departure from the engagement-based ranking paradigm of existing platforms. This possibility raises important questions for who gets to determine this information (platforms? users?) and how are these value profiles stored. We note that our method for personalizing value predictions and the corresponding VCQ do not rely on any privacy-compromising or sensitive individual attributes, and therefore representing a promising and procedurally legitimate approach to store value profiles. 

Furthermore, there is the concern that the value-based ranking approach outlined above could induce value-based filter bubbles. This could be mitigated by identifying and design for \textit{bridging values} that are held across partisan divides. However future work is needed to explore how what these bridging values are and how effective they are in mitigating value-based filter bubbles. 

We note that our approach relies on the OpenAI API which, if used in a real social media setting, could cause privacy concerns for users who have not explicitly consented to sharing their feeds’ content or values with a third-party service. This could be mitigated by local LLM implementations that do not share data with a third-party, or by sufficiently explicit data-sharing agreements with users. 

This work has been approved by the IRB of our institution to ensure adherence to ethical standards. In all the reproductions of social media posts from our dataset in this paper, we have reworded the post, blurred meta-data and replaced meta-data to preserve anonymity
while maintaining value saliency.
\bibliography{aaai2026}

\renewcommand{\thetable}{A\arabic{table}}
\setcounter{table}{0} 
\renewcommand{\thefigure}{A\arabic{figure}}
\setcounter{figure}{0} 

\appendix
\label{sec:reference_examples}

\section*{Appendix}
\section{Evaluating LLM's for data processing}
\subsection{NSFW Classifier}
To evaluate our classifier for the comparatively straightforward, unambiguous and sparse task of identifying content that is not safe for work (NSFW), we compute agreement on a random sample of 100 positive and 100 negative examples tagged by the classifier. Two authors (A1, A2) annotated the 200 posts with a binary label for NSFW based on the codebook shown in Table~\ref{tab:nsfw}. A1 and A2 agree with each other with Cohen's $\kappa=0.65$. A1 agreed with the model with Cohen's $\kappa=0.70$  corresponding to an accuracy of 84.9\%. A2 agreed with the model with Cohen's $\kappa=0.86$ corresponding to an accuracy of 93.0\%.

\subsection{Comprehensibility Classifier}
Next, we evaluate our classifier for the relatively ambiguous,  context-based and label balanced task of identifying content that is comprehensible and therefore capable of being judged for values. Comprehensibility is a highly subjective task because annotators are coming from different baselines, where what is comprehensible to them may differ based on their backgrounds, experiences and contexts. To evaluate this model, we also compute agreement on a random sample of 100 positive and 100 negative examples tagged by the classifier. Three authors (A1, A2, A3) annotated the 200 posts with a binary label for comprehensibility based on the codebook shown in Table~\ref{tab:comp}. We then computed an overall majority label from the three rater's responses (121 labeled as comprehensible, 79 labeled as incomprehensible) and downsample the comprehensible posts to balance the labels with respect to the human-rated majority labels. 

For this balanced set of annotations, the classifier has a recall of 0.695 and a precision of 0.721 (see Table~\ref{tab:comprehensibility_results}). However many of the misclassifications occur for posts where raters disagreed on if the post was comprehensible: when considering the subset of ratings where all three raters unanimously agreed on the label (85 labeled as comprehensible, 48 labeled as incomprehensible, similarly downsampling the comprehensible posts to balance the labels with respect to the ground truth), the classifier has a recall of 0.709 and a precision of 0.813. 

\begin{table}[ht]
\centering
\caption{Evaluation metrics for comprehensibility classifier grounded in 1) majority annotations from the three raters, and 2) majority annotations where the raters unanimously agreed.}
\label{tab:comprehensibility_results}
\begin{tabular}{lcc}
\hline
\textbf{Metric} & \textbf{Majority} & \textbf{Unanimous Majority} \\
\hline
Accuracy              & 0.703 & 0.740 \\
$\kappa$                 & 0.405 & 0.479 \\
Recall  & 0.695 & 0.709 \\
Precision             & 0.722 & 0.813 \\
F1 Score              & 0.708 & 0.757 \\
\hline
Number of posts              & 158 & 96 \\
\hline
\end{tabular}
\end{table}

\subsection{Preliminary Values Classifier}
To evaluate the preliminary values classifier we used to upsample infrequent values into the final annotation dataset, we leverage the fact that our annotated dataset contains examples of positive and negative labels of the preliminary values classifier for each of the 19 values. In particular, for each value $v$, we compare the mean human rating for posts we sampled as positive examples (i.e. the classifier identified as containing the value $\hat{y}_v\geq4$) to the other posts sampled for other values (i.e. $\hat{y}_v<4$), the results of which are shown in  Table~\ref{tab:pvc_eval}. We see that for 18/19 of the values, the average rating of the posts sampled via the preliminary values classifier is statistically greater than the those that were not. 
\begin{table}[ht]
\centering
\caption{Schwartz Value Rating Effects}
\label{tab:pvc_eval}
\begin{tabular}{lccc}
\hline
\textbf{Value} & \textbf{Avg. Rating} & \textbf{Avg. Rating} & \textbf{$p$}-val \\
 &   $(\hat{y}_v\geq4)$ & $(\hat{y}_v<4)$ &\\
\hline
Face & 1.174 & 0.761 & $< 0.001$ \\
Dominance & 0.472 & 0.371 & 0.001 \\
Resources & 0.844 & 0.390 & $< 0.001$ \\
Achievement & 1.245 & 0.517 & $< 0.001$ \\
Hedonism & 2.457 & 0.955 & $< 0.001$ \\
Self-Dir. Thoughts & 1.423 & 0.719 & $< 0.001$ \\
Self-Dir.Actions & 1.322 & 0.734 & $< 0.001$ \\
Stimulation & 1.238 & 0.583 & $< 0.001$ \\
Personal Security & 0.627 & 0.208 & $< 0.001$ \\
Societal Security & 1.389 & 0.649 & $< 0.001$ \\
Tradition & 1.677 & 0.683 & $< 0.001$ \\
Rule Conformity & 1.989 & 0.620 & $< 0.001$ \\
Interp. Conformity & 0.787 & 0.631 & $< 0.001$ \\
Humility & 1.426 & 1.345 & 0.171 \\
Dependability & 1.865 & 1.235 & $< 0.001$ \\
Caring & 2.551 & 1.235 & $< 0.001$ \\
Universal Concern & 2.446 & 0.851 & $< 0.001$ \\
Pres. of Nature & 3.076 & 0.465 & $< 0.001$ \\
Tolerance & 1.803 & 0.754 & $< 0.001$ \\
\hline
\end{tabular}
\end{table}

\section{Case studies of value disagreements in social media posts}
What kind of disagreements did we notice in social media posts? While we did not observe any global linguistic or contextual trends associated disagreements, we did observe many disagreements where certain value dimensions of a post where salient to some participants, while others dimensions were salient to others. In addition to annotating the posts for values, our survey also provided participants an optional space to provide their rationale. These rationales help shed light on some disagreements and divergent perspectives within the posts. 
\begin{figure}[t!]
\centering
\includegraphics[width=.99\linewidth]{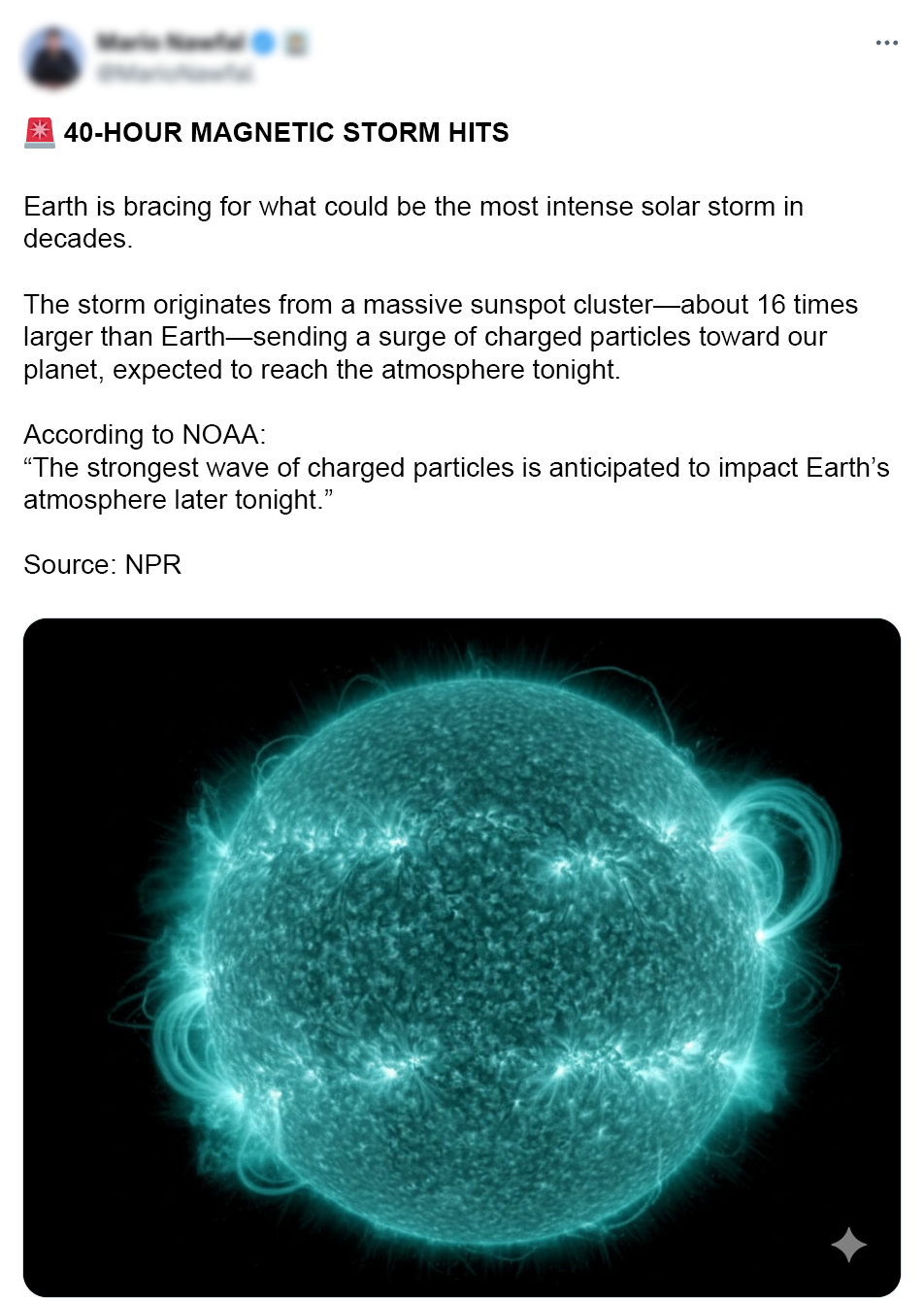}
\caption{Anonymized example post from our dataset.}
\label{fig:flare}
\end{figure}
\subsection{Case Study 1:}
Figure~\ref{fig:flare} depicts one post from our dataset, describing a magnetic storm hitting earth. On one hand, One participant read the post as a socially meaningful act---inferring care or universal concern from the decision to share timely information about a potentially impactful event, stating ``great post to give others information about world shifting weather events.'' On the other hand, another participant grounds value inference in the described phenomenon itself, focusing on the natural event’s aesthetic and reminder of humans' smallness in the face of nature, mapping it onto humility and appreciation of nature, stating ``The northern lights are just beautiful and a good reminder of the beauty of earth.''

\subsection{Case Study 2:}
\begin{figure}[h!]
\centering
\includegraphics[width=.99\linewidth]{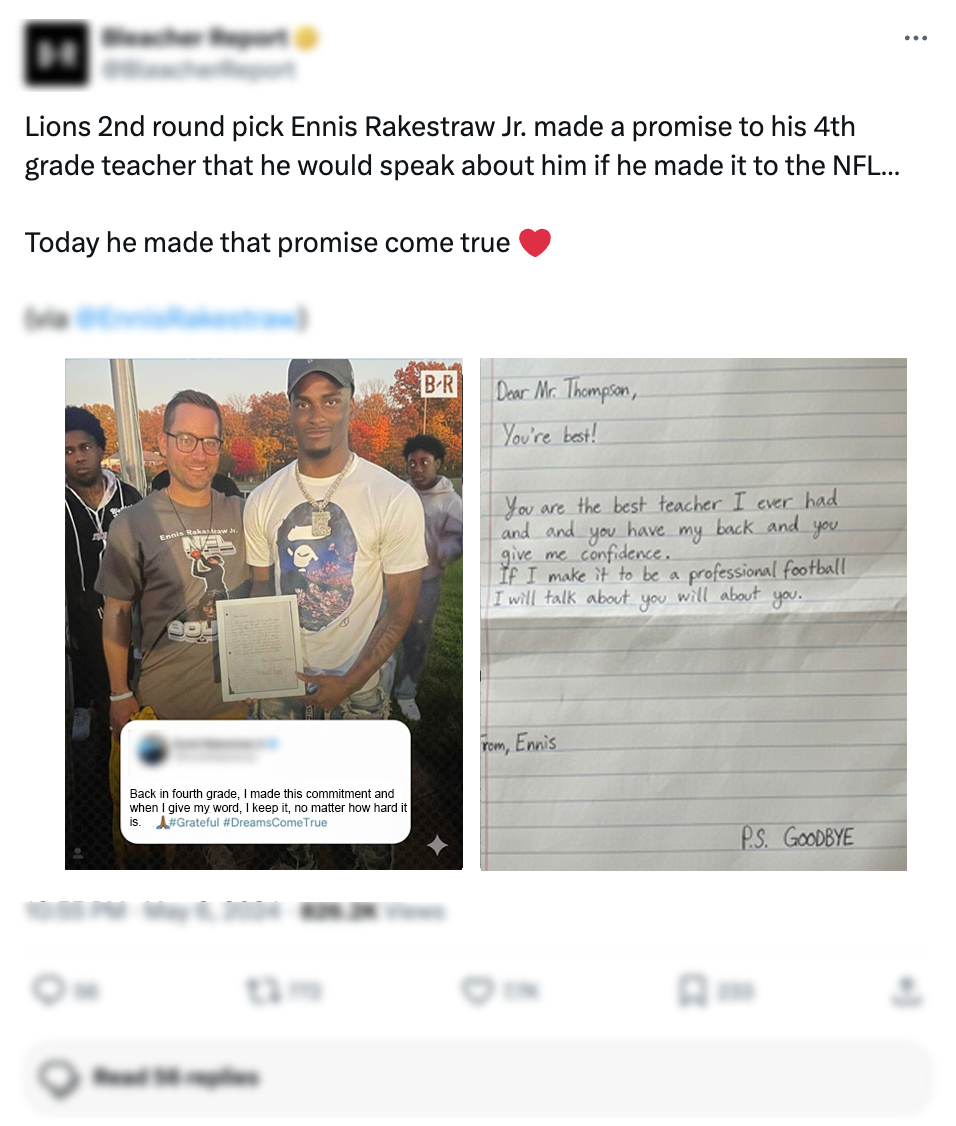}
\caption{Anonymized example post from our dataset. }
\label{fig:bleacher}
\end{figure}

Figure~\ref{fig:bleacher} depicts another post from our dataset, describing an NFL playing giving a shout out to their 4th grade teacher. One participant centers the relational and moral dimension of the story, interpreting the public acknowledgment of a teacher as an expression of dependability and caring toward those who supported the athlete’s development, stating: ``it’s important to thank those who inspire us''. The other focuses on the athlete’s success itself, interpreting the post as a story about achievement and public recognition, stating: ``It is a positive story about the guy who was drafted. It is about his success.''

\section{Attention checks}
Our first attention check included an image of the digits 15 and a textbox asking ``Please enter the number you see in the image (use numerical digits)'' Our second attention check asked the participant: ``Help us keep track of who is paying attention, please select -  "Somewhat disagree"  in the options below.'')

\begin{figure*}[b!]
\centering
\includegraphics[width=.99\linewidth]{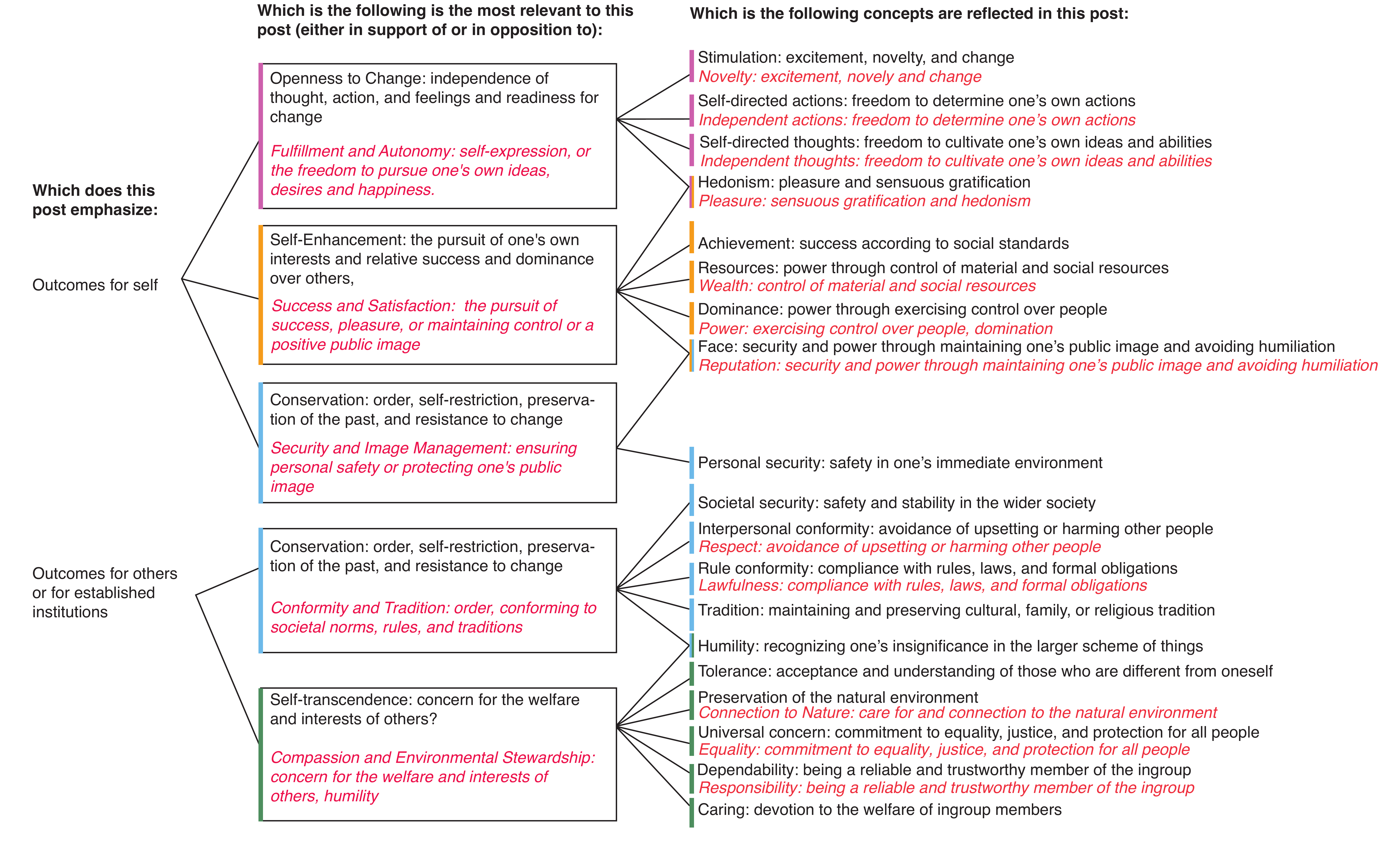}
\includegraphics[width=.99\linewidth]{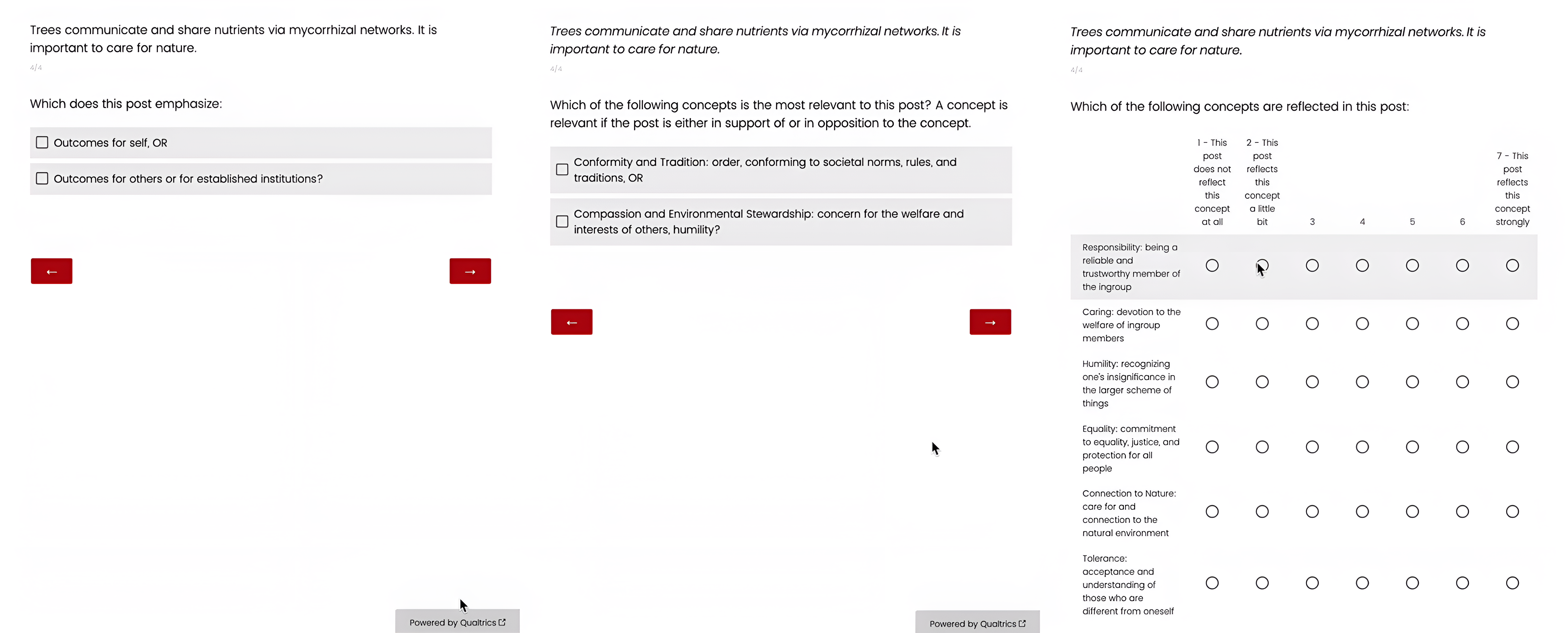}
\caption{Top: Recursive tree structure for value annotation. Red indicates modifications we took to make the categories more interpretable for raters and salient to the social media context. Bottom: Example of recursive labeling scheme for gating post}
\label{fig:tree}
\end{figure*}

\begin{figure*}[b!]
\centering
\includegraphics[width=.99\linewidth]{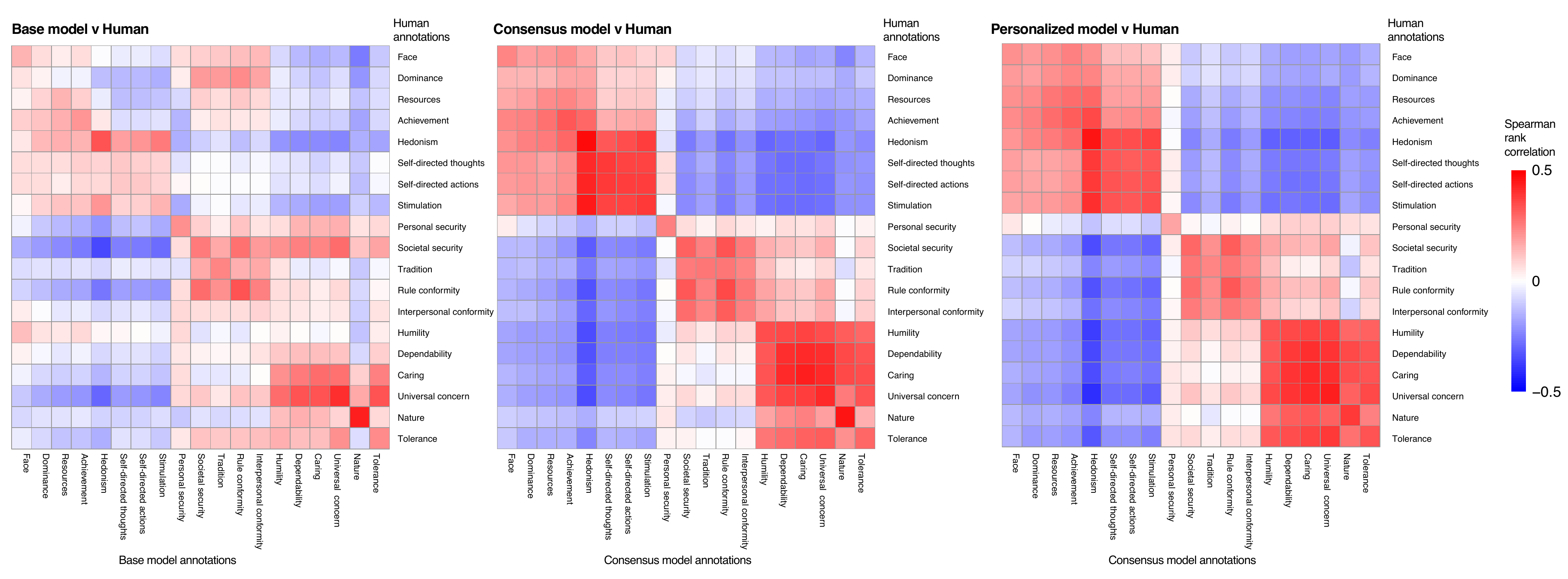}
\includegraphics[width=0.99\linewidth]{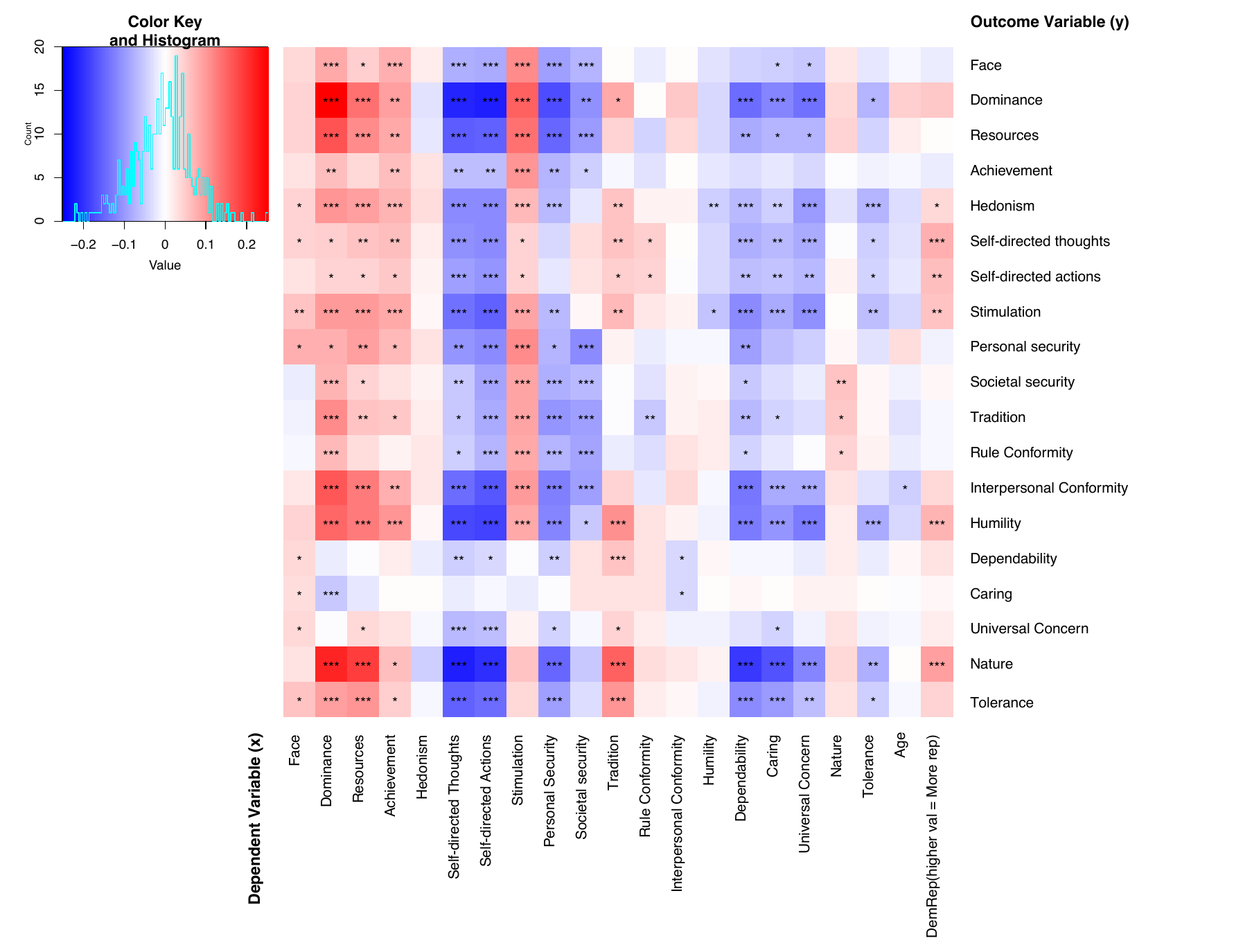}
\caption{Top: Value-wise spearman rank correlation between raters and base model (left), consensus model (center) and personalized model (right). Bottom: Regression coefficients in predicting value annotations (y) from a given value that that individual holds (x), as well as demographics such as age and partisanship.}
\label{fig:valueheatmap}
\end{figure*}

\begin{table*}[]
\centering
\scriptsize
\begin{tabular}{| c | p{0.65\linewidth} | p{0.25\linewidth} | }
\hline
\textbf{Index} & \textbf{Post Text} & \textbf{Value} \\ 
\hline \hline
1 & Pat Cummins decided to take a break from IPL 2023 due to hectic cricket season in International cricket then; - Won the WTC as captain. - Retained Ashes as captain. - Won the WC as captain. Captain, Leader, Legend, Cummins. & Achievement: success according to social standards \\\hline
2 & TOMORROW: I have an hourlong exclusive that we are broadcasting live for @BloombergTV with Citadel founder Ken Griffin, whose hedge fund recently surpassed Bridgewater as the world’s most profitable. Send questions!! There’s a lot on the agenda… & Achievement: success according to social standards \\\hline
3 & Being pro-palestine doesn't EVER mean that it gives you the right to be antisemitic. Jewish people have been our greatest allies, brothers, sisters since 7th october than anyone else on this fucking earth. if you're being antisemitic, youre not pro-palestine. we dont want you & Caring: devotion to the welfare of ingroup members \\\hline
4 & This morning, I proudly signed the Korean American VALOR Act into law – providing a pathway for thousands of Korean American Vietnam War veterans to access VA health care. Because of this law, they'll have the care and benefits they earned. & Caring: devotion to the welfare of ingroup members \\\hline
5 & Over 2000 years ago a child was born that came to die for all of our sins. He is the Christ, the living God, the Saviour of the world. Love Him with all your heart, because He loves you. Have a blessed Christmas everyone. & Caring: devotion to the welfare of ingroup members \\\hline
6 & R.I.P. My father, Marine Staff Sergeant Randolph Elder; my little brother Dennis Elder, Army Vietnam Vet; my older brother Kirk Randolph Elder, Navy Vietnam Era Vet; my nephew Eric Randolph Elder, Army Congressional Gold Medal: \#VeteransDay2023 & Caring: devotion to the welfare of ingroup members \\\hline
7 & The choice for GOP primary voters: Do we want “reform” or revolution? Do we aspire to “normalcy” or excellence? Do we want Super PAC puppets or patriots who speak the TRUTH? Clinton, IA . & Lawfulness: compliance with rules, laws, and formal obligations \\\hline
8 & After years, the People's Republic of China and the United States are restarting cooperation on counternarcotics. In particular, we seek to reduce the flow of precursor chemicals and pill presses fueling the fentanyl crisis. & Rule conformity: compliance with rules, laws, and formal obligations \\\hline
9 & A man has been arrested on suspicion of manslaughter following the death of ice hockey player Adam Johnson during a game in October, UK police say. & Personal security: safety in one’s immediate environment \\\hline
10 & getting the nice plates out for the holidays & Novelty: excitement, novelty, and change \\\hline
11 & getting the nice plates out for the holidays & Hedonism: pleasure and sensuous gratification \\\hline
12 & Darius Jackson allegedly denies abusing Keke Palmer. A “source” tells TMZ they were arguing over custody and photos in restraining order were him and Palmer wrestling over her phone. Also included is a video of Palmer’s mom threatening to put a bullet in his head :… Keke Palmer cradling her baby son Leodis as she steps out for the first time since being granted temporary sole custody and a restraining order against her allegedly 'abusive' ex Darius Jackson: & Reputation: security and power through maintaining one’s public image and avoiding \\\hline
13 & Merry Christmas to ALL my followers on I am GUILTY of loving America too much I am GUILTY of being ULTRA MAGA I am GUILTY of loving all of you Am I on the Naughty List ? & Resources: power through control of material and social resources \\\hline
14 & ``Every day is a gift.'' -- Art Loveley & Self-directed actions: freedom to determine one’s own actions \\\hline
15 & Couple breaks down when their cat returns after going missing & Stimulation: excitement, novelty, and change \\\hline
16 & Being pro-palestine doesn't EVER mean that it gives you the right to be antisemitic. Jewish people have been our greatest allies, brothers, sisters since 7th october than anyone else on this fucking earth. if you're being antisemitic, youre not pro-palestine. we dont want you & Tradition: maintaining and preserving cultural, family, or religious tradition \\\hline
17 & Over 2000 years ago a child was born that came to die for all of our sins. He is the Christ, the living God, the Saviour of the world. Love Him with all your heart, because He loves you. Have a blessed Christmas everyone. & Tradition: maintaining and preserving cultural, family, or religious tradition \\\hline
18 & The choice for GOP primary voters: Do we want “reform” or revolution? Do we aspire to “normalcy” or excellence? Do we want Super PAC puppets or patriots who speak the TRUTH? Clinton, IA . & Tradition: maintaining and preserving cultural, family, or religious tradition \\\hline
19 & UFC Champion Islam Makhachev just shared this pro-Palestine graphic! & Universal concern: commitment to equality, justice, and protection for all people \\\hline
20 & I'm proud to establish a new White House Initiative on Women’s Health Research, an effort led by the First Lady and my Gender Policy Council. Together, they'll work to ensure our Administration does everything it can to drive innovation in women’s health and close research gaps. & Equality: commitment to equality, justice, and protection for all people \\\hline
21 & Hey, remember when everyone endured like two straight years of trauma and then did nothing to address it and we all collectively and institutionally buried it way down deep under the constant productive pressures of capitalism? Probably fine & Equality: commitment to equality, justice, and protection for all people \\\hline
22 & Being pro-palestine doesn't EVER mean that it gives you the right to be antisemitic. Jewish people have been our greatest allies, brothers, sisters since 7th october than anyone else on this fucking earth. if you're being antisemitic, youre not pro-palestine. we dont want you & Equality: commitment to equality, justice, and protection for all people \\\hline
23 & Merry Christmas to ALL my followers on I am GUILTY of loving America too much I am GUILTY of being ULTRA MAGA I am GUILTY of loving all of you Am I on the Naughty List ? & Independent thinking: freedom to cultivate one’s own ideas and abilities \\\hline
24 & It's ironic to me that people who deny Jesus still observe Christmas and Easter by taking those days off. Devout atheists should work right through the holidays if they are serious about their denial. & Independent thinking: freedom to cultivate one’s own ideas and abilities \\\hline
25 & TOMORROW: I have an hourlong exclusive that we are broadcasting live for @BloombergTV with Citadel founder Ken Griffin, whose hedge fund recently surpassed Bridgewater as the world’s most profitable. Send questions!! There’s a lot on the agenda… & Wealth: control of material and social resources \\\hline
\end{tabular}
\caption{Value Calibration Questionnaire (VCQ) developed to personalize value predictions}
\label{tab:vcq}
\end{table*}

\begin{table*}
\begin{formal}
\tiny
\begin{textttenv}
\vspace{0.1cm}
Task -- I will provide a Twitter post alongside a codebook that describes in which settings Twitter posts are comprehensible Tweets, and in which settings Tweets lack understandability. I want you to apply each concept in the codebook to determine why and to what degree the concept the applies to the post. As you answer, please take the following steps:

Step 1) For each concept in the codebook, describe whether and to what degree the Tweet illustrates comprehensible, or uncomprehensible behavior, with the following format: {"<CONCEPT 1>": {"Why": "<explanation of how the concept applies to the post>", "Rating": <rating>}, ..., "<CONCEPT N>": {"Why": "<explanation of how the concept applies to the post>", "Rating": <rating>}}. ("Codebook Application")

Use the following scale to assign your rating:

0="the post strongly exhibits uncomprehensible behavior for the given concept"

1="the post somewhat exhibits uncomprehensible behavior for the given concept"

2="the post somewhat exhibits comprehensible behavior for the given concept"

3="the post strongly exhibits comprehensible behavior for the given concept" 

Step 2) Summarizing your reasoning in Steps 1 and 2, determine a single rating for whether the post is comprehensible, or uncomprehensible: "Final Rating": {"Why": "<explanation of final rating>", "Rating": <rating>}. ("Agreement Rating")

Use the following scale to assign your rating:
0="the post strongly exhibits uncomprehensible behavior"
1="the post somewhat exhibits uncomprehensible behavior"
2="the post somewhat exhibits comprehensible behavior"
3="the post strongly exhibits comprehensible behavior"

Codebook --

READABILITY: Readable -- Well-structured, coherent, and easily understandable language that facilitates the comprehension of the expressed values;Unreadable -- Poorly structured, incoherent, or overly complex language that hinders understanding of the message

COHERENCE: Coherent -- Logically structured, with a clear progression of ideas that underscore the expressed meaning; Incoherent -- Disjointed or lacking logical flow, making it difficult to discern any underlying meaning

SPAM BEHAVIOR: Authentic -- Content is original, personal, and specifically targets relevant issues; Spammy -- Repetitive, unsolicited, or irrelevant content or appears automated

CONTEXT REQUIRED FOR UNDERSTANDING: SelfContained -- Minimal to no additional context required, as the post is self-contained and clear; RequiresContext -- Requires external information or additional context for understanding due to vagueness or lack of focus

Output Format -- Output your message in JSON, where you provide the following:
{
"Codebook Application": {        
"<CONCEPT 1>": {
"Why": "<explanation of how the concept applies to the post>", "Rating": <integer from 0-3>},         ...,
"<CONCEPT N>": {            "Why": "<explanation of how the concept applies to the post>", "Rating": <integer from 0-3>}
},
"Final Rating": {            "Why": "<explanation of final rating>", "Rating": <integer from 0-3>}
}

Social Media Post -- 

\#\#\#
\{post\}
\#\#\#
\end{textttenv}
\end{formal}
\caption{Prompt for comprehension check}
\label{tab:comp}
\end{table*}

\begin{table*}
\begin{formal}
\tiny
\begin{textttenv}
\vspace{0.1cm}
Task -- I will provide a Twitter post alongside a codebook that describes in which settings Twitter posts are comprehensible Tweets, and in which settings Tweets lack understandability. I want you to apply each concept in the codebook to determine why and to what degree the concept the applies to the post. As you answer, please take the following steps:

Step 1) For each concept in the codebook, describe whether and to what degree the Tweet illustrates comprehensible, or uncomprehensible behavior, with the following format: {"<CONCEPT 1>": {"Why": "<explanation of how the concept applies to the post>", "Rating": <rating>}, ..., "<CONCEPT N>": {"Why": "<explanation of how the concept applies to the post>", "Rating": <rating>}}. ("Codebook Application")

Use the following scale to assign your rating:

    0="This post does not contain NSFW content at all"
    
    1="This post contains a little bit of NSFW content"
    
    2="This post moderately contains NSFW content"
    
    3="This post strongly contains NSFW content"
    
Step 2) Summarizing your reasoning in Steps 1 and 2, determine a single rating for whether the post is NSFW: "Final Rating": {"Why": "<explanation of final rating>", "Rating": <rating>}. ("Agreement Rating")

Use the following scale to assign your rating:

    0="This post does not contain NSFW content at all"
    
    1="This post contains a little bit of NSFW content"
    
    2="This post moderately contains NSFW content"
    
    3="This post strongly contains NSFW content"

Codebook --

SEXUAL: Sexual -- Contains graphic imagery or text that is sexually explicit; NotSexual -- Does not contain sexually explicit imagery. Safe for the workplace.

VIOLENT: Violent -- Contains graphic imagery or text that describes violence in an explicit or egregieous way; NonViolnet -- Contains no reference to violence

DEROGATORY: Derogatory -- Contains derogatory imagery or text that could make someone uncomfortable; NonDerogatory -- Does not contain any derogatory imagery or text

Output Format -- Output your message in JSON, where you provide the following:

{
"Codebook Application": {        
"<CONCEPT 1>": {
"Why": "<explanation of how the concept applies to the post>", "Rating": <integer from 0-3>},         ...,
"<CONCEPT N>": {            "Why": "<explanation of how the concept applies to the post>", "Rating": <integer from 0-3>}
},
"Final Rating": {            "Why": "<explanation of final rating>", "Rating": <integer from 0-3>}
}

Social Media Post -- 

\#\#\#
\{post\}
\#\#\#
\end{textttenv}
\end{formal}
\caption{Prompt for NSFW check}
\label{tab:nsfw}
\end{table*}

\begin{table*}
\begin{formal}
\small
\begin{textttenv}
\vspace{0.1cm}
'Consider the following set of concepts, listed as !<CONCEPT>! : !<DEFINITION>!

- FACE\_SCHWARTZ: Security and power through maintaining one's public image and avoiding humiliation; not wanting to be shamed by anyone, protecting one's public image, wanting to always be treated with respect and dignity by people

- DOMINANCE\_SCHWARTZ: Power through exercising control over people; wanting people to do what one says, wanting to be the most influential person in any group, wanting to be the one who tells others what to do

- RESOURCES\_SCHWARTZ: Power through control of material and social resources; wanting to have the feeling of power that money can bring, wanting to be wealthy, pursuit of high status and power

- ACHIEVEMENT\_SCHWARTZ: Success according to social standards; being ambitious, wanting to be successful, wanting people to admire one\'s achievements

- HEDONISM\_SCHWARTZ: Pleasure and sensuous gratification; having a good time, enjoying life\'s pleasures, taking advantage of every opportunity to have fun

- SELF\_DIRECTED\_THOUGHTS\_SCHWARTZ: Freedom to cultivate one's own ideas and abilities; being creative, forming one's own opinions and having original ideas, learning things for oneself and improving one's abilities

- SELF\_DIRECTED\_ACTIONS\_SCHWARTZ: Freedom to determine one's own actions; making one\'s own decisions about one's life, doing everything independently, freedom to choose what one does

- STIMULATION\_SCHWARTZ: Excitement, novelty, and change; looking for different kinds of things to do, excitement in life, wanting to have all sorts of new experiences

- PERSONAL\_SECURITY\_SCHWARTZ: Safety in one's immediate environment; avoiding anything that might endanger one\'s safety, personal security, wanting to live in secure surroundings

- SOCIETAL\_SECURITY\_SCHWARTZ: Safety and stability in the wider society; wanting one\'s country to protect itself against all threats, wanting the state to be strong so it can defend its citizens, wanting to have order and stability in society

- TRADITION\_SCHWARTZ: Maintaining and preserving cultural, family, or religious traditions; maintaining traditional values's or beliefs, following one's family customs or the customs of a religion, valuing the traditional practices of one's culture

- RULES\_CONFORMITY\_SCHWARTZ: Compliance with rules, laws, and formal obligations; wanting to do what people in authority say, following rules even when no one is watching, obeying all the laws

- INTERPERSONAL\_CONFORMITY\_SCHWARTZ: Avoidance of upsetting or harming other people; avoiding upsetting other people, not being annoying to anyone, trying to be tactful and avoid irritating people

- HUMILITY\_SCHWARTZ: Recognizing one's insignificance in the larger scheme of things; not drawing attention to oneself, being humble, being satisfied with what one has and not to ask for more

- DEPENDABILITY\_SCHWARTZ: Being a reliable and trustworthy member of the ingroup; being loyal to those who are close, going out of one's way to be a dependable and trustworthy friend, wanting those one spends time with to be able to rely on one completely

- CARING\_SCHWARTZ: Devotion to the welfare of ingroup members; helping people who are dear to oneself, caring for the well-being of people one is close to, trying to always be responsive to the needs of family and friends

- UNIVERSAL\_CONCERN\_SCHWARTZ: Commitment to equality, justice, and protection for all people; protecting society's weak and vulnerable members, wanting every person in the world to have equal opportunities in life, wanting everyone to be treated justly even people one doesn't know

- PRESERVATION\_OF\_NATURE\_SCHWARTZ: Preservation of the natural environment; believing that one should care for nature, working against threats to the world of nature, protecting the natural environment from destruction or pollution

- TOLERANCE\_SCHWARTZ: Acceptance and understanding of those who are different from oneself; promoting harmony and peace among diverse groups, listening to people who are different from oneself, wanting to understand people even when disagreeing with them

For each concept, rate whether the message reflects the concept on a Likert scale with options [0, 1, 2, 3, 4, 5, 6], where:
0 = This post does not reflect this concept at all
1 = This post reflects this concept a little bit 
6 = This post reflects this concept strongly

A concept can be reflected if the message either supports or opposes the concept.

Output: One JSON dictionary of the following format. Rating is a dictionary of key value pairs, with each key being a concept and each value your rating for that concept:

{ "Rating": {$\backslash$'Concept$\backslash$': !<RATING>!} }

\end{textttenv}
\end{formal}
\caption{Prompt for value expression}
\label{tab:values}
\end{table*}

\begin{figure*}[h]
\centering
\includegraphics[width=.79\linewidth]{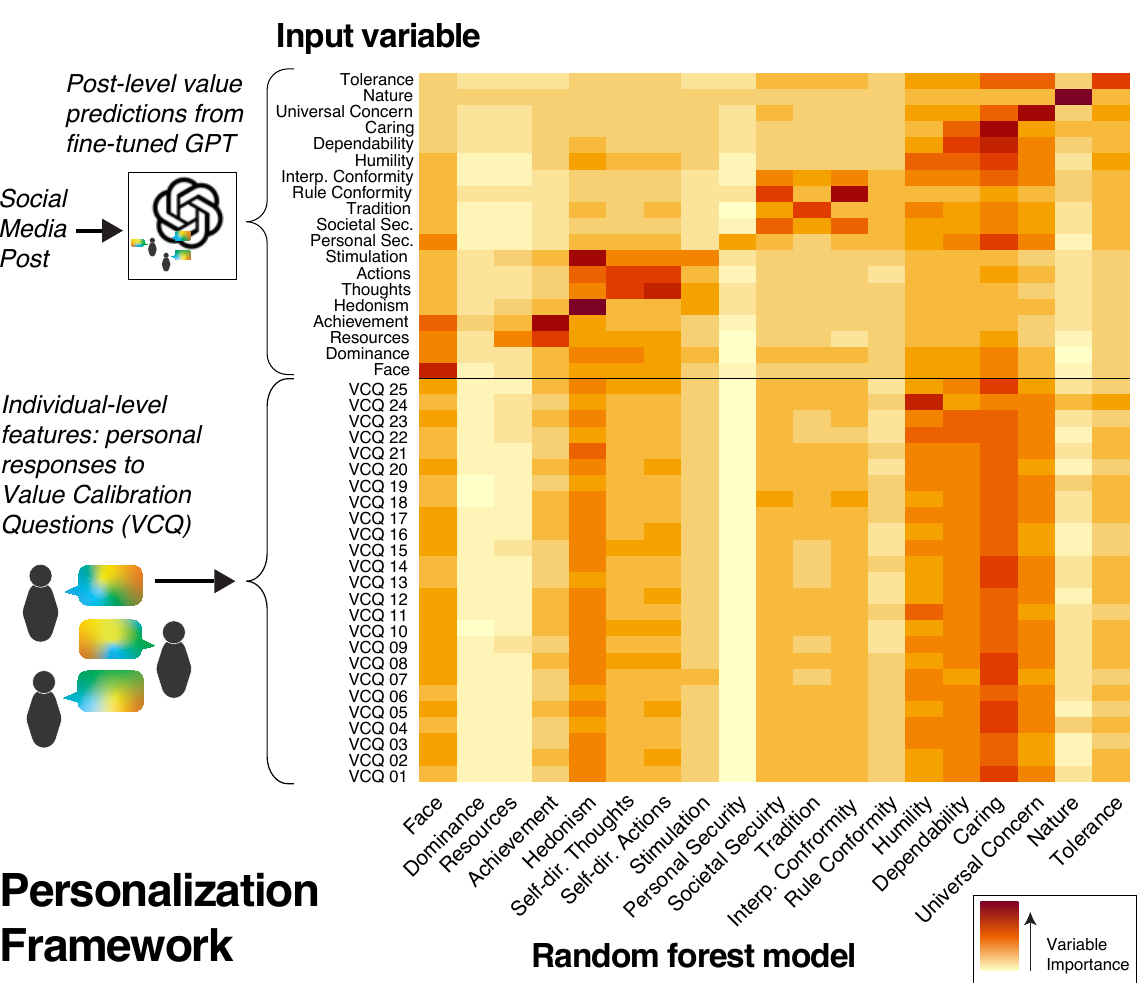}
\caption{Heatmap of variable importance for the 19 random forest
models (measured by total decrease in node impurities).}
\label{fig:rfflow_sup}
\end{figure*}

\end{document}